\documentclass[apjl]{emulateapj}
\usepackage{natbib}
\usepackage{float}
\usepackage{epsfig}
\usepackage{amssymb}
\usepackage{amsmath}

\newcommand{\bn}{\begin{enumerate}}
\newcommand{\en}{\end{enumerate}}
\newcommand{\bi}{\begin{itemize}}
\newcommand{\ei}{\end{itemize}}

\newcommand{\Msun}{M_\odot}

\newcommand{\himpc}{h^{-1} {\rm Mpc}}
\newcommand{\hikpc}{h^{-1} {\rm kpc}}

\newcommand{\muv}{\rm M_{uv}}
\newcommand{\mpct}{{\rm Mpc}^{-3}}
\newcommand{\myr}{\rm Myr}

\newcommand{\Htwo}{\rm H_2}

\shorttitle{Connecting the Dots}
\shortauthors{Jaacks et al.}

\begin{document}
\title{Connecting the Dots: Tracking Galaxy Evolution Using Constant Cumulative Number Density at $3\leq \lowercase{z} \leq 7$}

\author{Jason Jaacks$^1*$, Steven L. Finkelstein$^1$ \& Kentaro Nagamine$^{2,3}$}
\affil{$^1$Department of Astronomy, The University of Texas at Austin, Austin, TX 78712 \\
$^2$Department of Earth and Space Science, Graduate School of Science, Osaka University,\\
1-1 Machikaneyama-cho, Toyonaka, Osaka 560-0043, Japan\\
$^3$Department of Physics \& Astronomy, University of Nevada, Las Vegas,\\ 4505 S. Maryland Pkwy, Las Vegas, NV, 89154-4002, USA}
\email{* jaacks@astro.as.utexas.edu}


\begin{abstract}
Using the cosmological smoothed particle hydrodynamical code {\small GADGET-3} we make a realistic assessment of the technique of using constant cumulative number density as a tracer of galaxy evolution at high redshift.  We find that over a redshift range of $3\leq z \leq7$ one can {\it on average} track the growth of the stellar mass of a population of galaxies selected from the same cumulative number density bin to within $\sim 0.20$ dex.  Over the stellar mass range we probe ($10^{10.39}\leq M_s/\Msun \leq 10^{10.75}$ at $z =$ 3 and $10^{8.48}\leq M_s/\Msun \leq 10^{9.55}$ at $z =$ 7) one can reduce this bias by selecting galaxies based on an evolving cumulative number density.  We find the cumulative number density evolution exhibits a trend towards higher values which can be quantified by simple linear formulations going as $-0.10\Delta z$ for descendants and $0.12\Delta z$ for progenitors.  Utilizing such an evolving cumulative number density increases the accuracy of descendant/progenitor tracking by a factor of $\sim2$.  This result is in excellent agreement, within $0.10$ dex, with abundance matching results over the same redshift range.  However, we find that our more realistic cosmological hydrodynamic simulations produce a much larger scatter in descendant/progenitor stellar masses than previous studies, particularly when tracking progenitors.  This large scatter makes the application of either the constant cumulative number density or evolving cumulative number density technique limited to average stellar masses of populations only, as the diverse mass assembly histories caused by stochastic physical processes such as gas accretion, mergers, and star formation of individual galaxies will lead to a larger scatter in other physical properties such as metallicity and star-formation rate.

\end{abstract}



\section{Introduction}
\label{sec:intro}

Understanding how galaxies evolved from minute perturbations in the distant Universe into the diverse zoo of shapes and sizes we see today is one of the fundamental goals of modern astronomy.  The current frontier lies at the edge of the observable universe, $\sim500\ \myr$ after the Big Bang, and is primarily possible using the Wide Field Camera 3 (WFC3) instrument aboard the {\it Hubble Space Telescope} ({\it HST}).  Using the Lyman-break technique and/or photometric redshifts to select candidate galaxies, programs such as CANDELS (PIs Faber \& Ferguson), BoRG \citep{Trenti.etal:11}, and HUDF09/UDF12 \citep{Bouwens.etal:11a,Ellis:13}, are able to identify galaxies down to rest-frame UV magnitudes of $\muv \sim-17.5$ \citep[e.g.,][]{Finkelstein.etal:12,Finkelstein:14, Trenti.etal:11, Bouwens.etal:12b,Ellis:13}.  Two of these galaxies have been spectroscopically confirmed to be the earliest known galaxies to date with redshifts of $z$=7.51 \citep{Finkelstein:13} and $z$=7.73 \citep{Oesch:15}.  With the help of the gravitational lensing effect of massive foreground galaxy clusters, campaigns such as CLASH \citep{Bouwens:14a} and the {\it HST Frontier Fields} will extend our understanding even further to $z\ge9$ and UV magnitudes as faint as $\muv \sim-13$, greatly increasing the dynamic range of the observed galaxy population for study.

From these surveys, fundamental properties such as stellar mass, age and star formation rate (SFR) can be derived by comparing the spectral energy distributions (SEDs) of galaxies to stellar population models.
By selecting galaxies at different epochs (``snapshots'' in time), we can in principle directly observe how galaxies evolve.  However, tracing galaxies from one epoch to another has proven challenging, and frequently accompanied by misinterpretation.
Previous studies have matched galaxies at different epochs by comparing samples selected to have similar physical tracers, such as at fixed UV luminosity \citep[e.g.,][]{Stark.etal:09}.  This can be problematic as two galaxies, one at $z$=6 and the other at $z$=3, with similar UV luminosities could have dramatically different mass assembly histories depending on their individual star formation histories (SFHs), environments and/or merger histories.  

A novel approach to this problem was suggested by \citet{vandokkum:10} who tracked a population of galaxies through cosmic time ($z$=2 to $z$=0.1) selected to have a constant number density, using the observed cumulative galaxy stellar mass function (CSMF).  The critical assumption this approach makes is that galaxies in the same number density bin will grow at a similar, smooth rate with a conserved rank order.  Monte Carlo simulations were utilized to test the effects of mergers and starbursts, and it was concluded that only 1:1 mass-ratio mergers were able to break the rank order of galaxies.  

The use of this technique was tested numerically by \citet{leja:13} using the Millennium dark matter (DM) halo catalog \citep{Springel:05b} and semi-analytic models \citep{Guo:11}.  They found that the actual mass growth tracked the mass growth inferred at a constant number density to within $40\%$ over a redshift range of $z$=3 to $z$=0.  They note, however, that they did not reproduce the observed mass evolution of galaxies.  This was attributed to the fact that the sub-grid physics (star formation, feedback, mergers) in the semi-analytic models  utilized for their project might not be accurately capturing the physics of the observed galaxies.

\citet{Papovich.etal:11} extended this technique to track the growth of observed galaxies at much higher redshift, from $z$=8 to $z$=3, concluding that galaxies exhibit a SFH that \emph{increases} with time (contrary to previous assumptions of decaying SFHs, see also \citet{Reddy:12,Finlator.etal:11,Jaacks.etal:12b}).
This result has critical implications for the estimates of physical properties of observed galaxies.  To justify the use of this method at high-redshift, the authors used the DM halo catalogs from the Millennium simulations to track halo growth, making the assumption that halo growth directly tracks galaxy growth (i.e., $n_{\rm gal}(>L_{\rm uv})\approx n_{\rm gal}(>M_{s})\approx n_{\rm gal}(>M_{\rm vir})$).  While they found that halo mass growth at a constant number density of $n=2\times10^{-4}\ \mpct$ was indeed smooth over the same time period, there was no consideration for the physics of formation and evolution of the galaxies themselves. 

A more recent numerical test at high-$z$ was performed by \cite{Behroozi:13} using the Bolshoi DM halo catalog \citep{Klypin:11} along with the merger tree and an abundance matching technique which assigns galaxies to halos based on their relative number densities.  This study found that when comparing stellar mass evolution across redshift one must account for the fact the the number density of a given population is not constant to avoid mass estimate errors of $\sim0.3$ dex.  While the technique used by \citet{Behroozi:13} did extend to tracking descendants of $z$=6 galaxies, it did not explicitly include the details of fundamental physics involved in galaxy formation and evolution. 

A contemporaneous work by \citet{Torrey:15} utilized the Illustris \citep{Vogelsberger:14} hydrodynamical simulations to test the constant number density evolution tracking method at $0\leq z \leq 3$, and found that galaxies do not evolve at constant number densities due to the effects of merger events and scatter in individual galaxies growth rates.  We will show that these results are very consistent to this work even though focused on different redshift epochs.

In this work we will use cosmological smoothed particle hydrodynamical code {\small GADGET-3}, where galaxies form in-situ, to test the technique of using constant cumulative number density as a tracer for tracking galaxy evolution at high-redshift (i.e. $3\leq z \leq 7$).  Motivated by recent tests of this technique which rely the combination of dark-matter-only simulations with either abundance matching or semi-analytic models, we intend to capture the realistic mass assembly histories of galaxies in simulations with sophisticated baryonic physics.  This work will be organized as follows: In Section~\ref{sec:sim} we will detail the simulations used in this exercise; Section~\ref{sec:methods} discusses the methods used for identifying and tracking galaxies in our simulations; Section~\ref{sec:results} presents our results; In Section~\ref{sec:disc} we present additional discussion, and we summarize in Section~\ref{sec:sum}. 


\section{Simulations}
\label{sec:sim}

\begin{table} 
\caption{Simulation Parameters}
\begin{tabular}{ccccc}
\hline
 Box size & $N_{p}$ & $m_{DM}$ & $m_{\rm gas}$ & $\epsilon$  \\ 
($\himpc$) & (DM, Gas)  & ($h^{-1} \Msun$) & ($h^{-1} \Msun$) & ($\hikpc$) \\
\hline\\ [-1.5ex]
$100.0$ & $2 {\times} 800^{3}$ & $1.17 {\times} 10^{8}$ & $2.38 {\times} 10^{7}$ & $5.00$  \\ [+1.0ex]
\hline \\ \vspace{-0.5cm}
\end{tabular}
Simulation parameters used in this paper. The parameter $N_p$ is the number of gas and dark matter particles; $m_{\rm DM}$ and $m_{\rm gas}$ are the particle masses of dark matter and gas; $\epsilon$ is the comoving gravitational softening length.
\label{tbl:Sim}
\end{table}

We use a modified version of the smoothed particle hydrodynamics (SPH) code GADGET-3 \citep[originally described in][]{Springel:05}.  Our code includes radiative cooling by H, He, and metals \citep{Choi:09}, heating by a uniform UV background \citep[UVB;][]{Faucher.etal:09}, the UVB self-shielding effect \citep{Nagamine.etal:10}, the initial power spectrum of \citet{Eisenstein&Hu:99}, supernovae feedback, a sub-resolution multiphase ISM model \citep{Springel:03}, Multicomponent Variable Velocity (MVV) wind model \citep{Choi:11a} and $\Htwo$ regulated star formation \citep{Thompson:14}.  Our current simulations do not include AGN feedback which is expected to contribute significantly at $z<3$ in massive galaxies.  See \citet{Thompson:14} and \citet{Jaacks:13} for detailed comparisons of our simulations to both high and low redshift observations.   In particular we find excellent agreement with both the observed $z$=4-7 UV luminosity functions in both faint-end slope ($\alpha$) and characteristic brightness ($M_{UV}^*$) found in \citet{Finkelstein:14} and \citet{Bouwens:15a}, as well as the number densities ($\phi$) from the observed stellar mass functions in \citet{Song:15} over the mass range studied here. 

\begin{table*} 
\begin{center}
\caption{Identification, mass and total number of galaxies for each number density bin used in this study}
\begin{tabular}{cccccc}
\hline
 ID &$n$ (center) & $\log n$ & $\log M_s$ at $z$=7  & $\log M_s$ at $z$=3	   & total \#  \\ 
    &  [Mpc$^{-3}$]  &          & [$\Msun$] 		   &   [$\Msun$]       		& \\
\hline\\ [-1.5ex]

n1    & $2.00\times10^{-4}$ & $-3.70\pm0.20$ & $8.70^{+0.22}_{-0.23}$ &$10.38^{+0.10}_{-0.10}$ & $538$\vspace{0.2cm}  \\
n2    & $8.10\times10^{-5}$ & $-4.10\pm0.20$ & $9.10^{+0.15}_{-0.18}$ &$10.54^{+0.08}_{-0.07}$ & $201$\vspace{0.2cm}  \\
n3    & $2.87\times10^{-5}$ & $-4.50\pm0.20$ & $9.40^{+0.15}_{-0.13}$ &$10.69^{+0.06}_{-0.05}$ & $75$\vspace{0.2cm}  \\
\hline 
\end{tabular}

\label{tbl:data}
\end{center}
\end{table*}

Our simulation is set up with $800^3$ particles for both gas and dark matter in a comoving box with sides of $100h^{-1}$ Mpc.  We will refer to this run as N800L100 and complete simulation parameters are summarized in Table~\ref{tbl:Sim}.  We use the 2LPTic \citep{Scoccimarro.etal:98} code which utilizes second-order Lagrangian perturbation theory for generating our initial conditions at $z$=99.  The adopted cosmological parameters are consistent with the WMAP7 data \citep{Komatsu:09}: ${\Omega_{\rm m}}=0.26$, ${\Omega_{\Lambda}}=0.74$, ${\Omega_{\rm b}}=0.044$, $h=0.72$, $\sigma_{8}=0.80$, $n_{s} =0.96$.

Snapshots were recorded every $20\ \myr$ starting at $z$=10 allowing us to more accurately track the growth of each galaxy, resulting in $72$ snapshots from $z$=7 to 3.  Galaxies were identified in each snapshot using a variant of the SUBFIND algorithm \citep{Springel:01} which groups particles based on baryonic density peaks.  A minimum of 32 particles (gas$+$stars) is required to be considered for grouping.  However we require a minimum of $100$ baryonic particles and at least 20 star particles for the galaxies to be considered within our mass resolution.  

It should be noted that our simulations assumed a Chabrier initial mass function (IMF) \citep{Chabrier:03} and when appropriate, data such as stellar masses which assumed a different IMF (e.g., \citet{Salpeter:55}) are converted for consistency using $M_{IMF}=M_{\rm Salpeter}/f_{IMF}$ where $f_{IMF}=1.6$ for Chabrier. \citep{Nagamine.etal:06,Raue.etal:12}.


\section{Methods}	
\label{sec:methods}
\subsection{Galaxy selection}
\label{subsec:select}

\begin{figure}
\centerline{\includegraphics[width=1.10\columnwidth,angle=0] {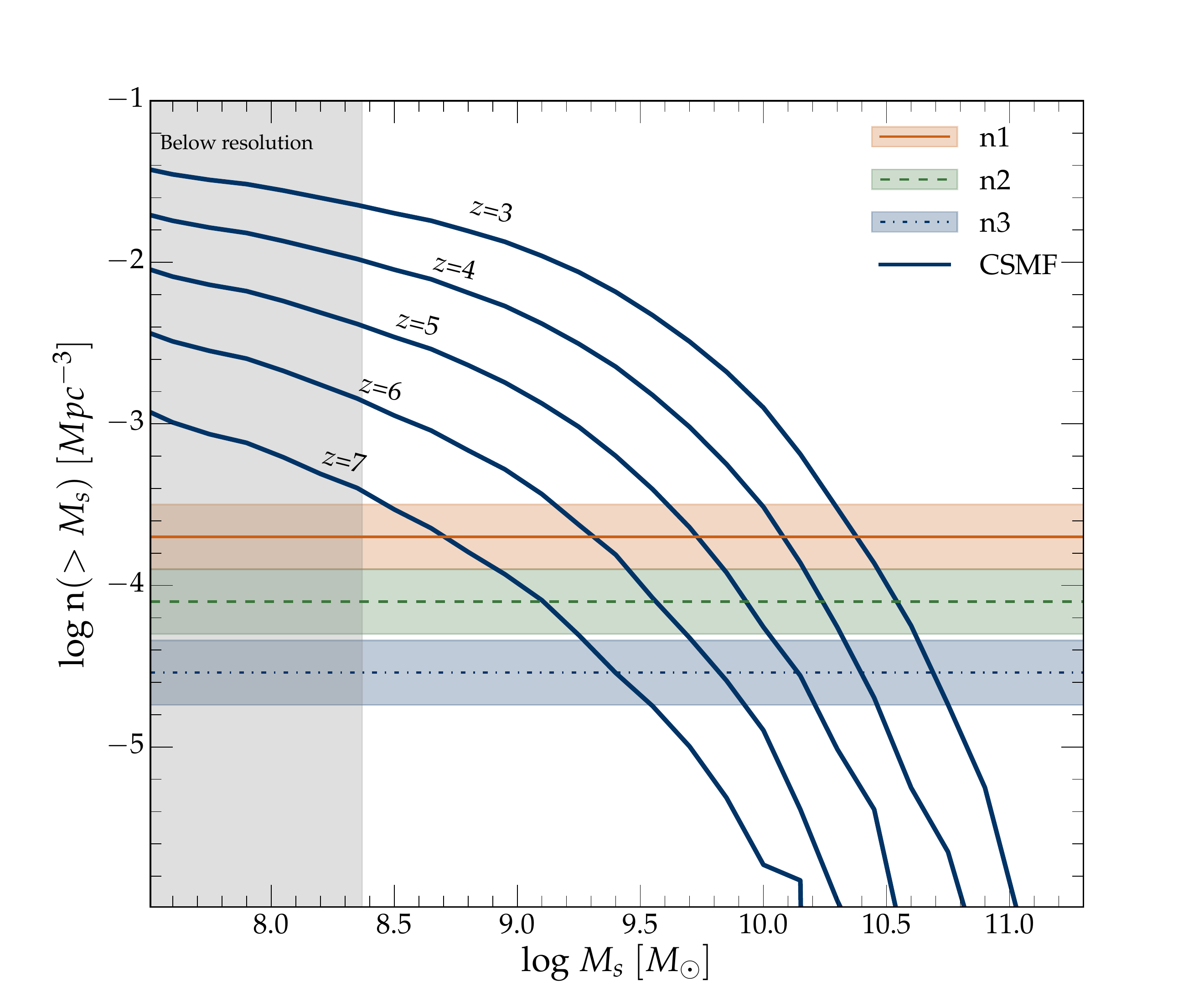}}
\caption{Cumulative stellar mass function for $z=3, 4, 5, 6\ {\rm and}\ 7$ (solid blue lines, top to bottom).  The horizontal solid, dashed and dashed-dot lines represent constant number density cuts n1, n2, n3 (see Table~\ref{tbl:data} for details), respectively.  For each number density studied, the total bin width is shown by the horizontal orange, green and blue shaded areas.  The gray shaded region denotes the approximate mass resolution for our simulation run, which corresponds to $\sim 100$ baryon particles.  The mass value associated with the intersection of the CSMF and the constant number density cuts will be referred to as the {\it inferred} mass growth.}
\label{fig:cmf}
\end{figure}

Galaxies are selected based on symmetric number density bins as determined by the cumulative stellar mass function (CSMF, Figure~\ref{fig:cmf}) from our simulation at $z$=7.  For our fiducial number density bin to select galaxies, we use $n=2.0\times10^{-4}$ [Mpc$^{-3}$], a choice which is motivated by its use in previous studies \citep[e.g.,][]{vandokkum:10,Papovich.etal:11,leja:13}.  This number density can then be translated via the CSMF to a corresponding stellar mass of $M_s\approx 10^{8.7} \Msun$ at $z$=7.  We also explore lower number densities of $n=8.10\times10^{-5}$ [Mpc$^{-3}$] and $n=2.87\times10^{-5}$ [Mpc$^{-3}$] which have corresponding $z =$ 7 masses of  $M_s\approx 10^{9.1} \Msun$ and $M_s\approx 10^{9.4} \Msun$, respectively.  Throughout this work these number density bins will be referred to as n1, n2, and n3, respectively. Full details of each number density bin studied can be found in Table~\ref{tbl:data}. The choices of the values for n2 and n3 are motived by covering a wide range of masses within our simulation mass resolution, with little to no overlap between populations.  We do not explore higher number density (lower mass) values due to the restrictions set by our simulation's resolution (see gray shading in Figure~\ref{fig:cmf}).  We also do not explore values lower than n3, as such a bin would have $<50$ tracked galaxies, leading to poor number statistics.

It should be noted that the symmetric number density bin leads to a slightly asymmetric mass bin (i.e. more low mass galaxies), depending on the shape of the CSMF at the selection redshift.  However this choice allows for a constant number of galaxies at each redshift for a given number density, whereas a symmetric mass bin would lead to varying galaxy counts due to the changing shape of the CSMF.

\subsection{Galaxy tracking}
\label{sec:tracking}
 The tracking of single galaxies across multiple snapshots is achieved through particle ID matching and the ``most massive" descendant/progenitor algorithm.  The galaxy at each subsequent snapshot which contains the highest number of matching particles from the original galaxy is considered to be its descendant/progenitor.


One downside to the ``most massive'' descendant/progenitor algorithm is that there are scenarios when the procedure is not completely reversible. For example, in merger events the progenitor path will choose the most massive of the two galaxies whereas the descendant path may have originated via the lower mass galaxy.  To test the impact of this effect on our results we conducted a test with a random sample population of galaxies selected at $z$=7.  This population was tracked to $z$=3 and the final $GIDs$ were used as a starting point for a progenitor track back to $z$=7.  We found that $\sim10\%$ of the progenitor population differed from the original $z$=7 population.  However, this difference only led to a median mass differential of $\sim0.02$ dex.  Therefore, we conclude that the partial non-reversibility of the ``most massive'' algorithm utilized in this work has a negligible impact on our results.

\begin{figure*}
\begin{center}
\includegraphics[width=1.4\columnwidth,angle=0] {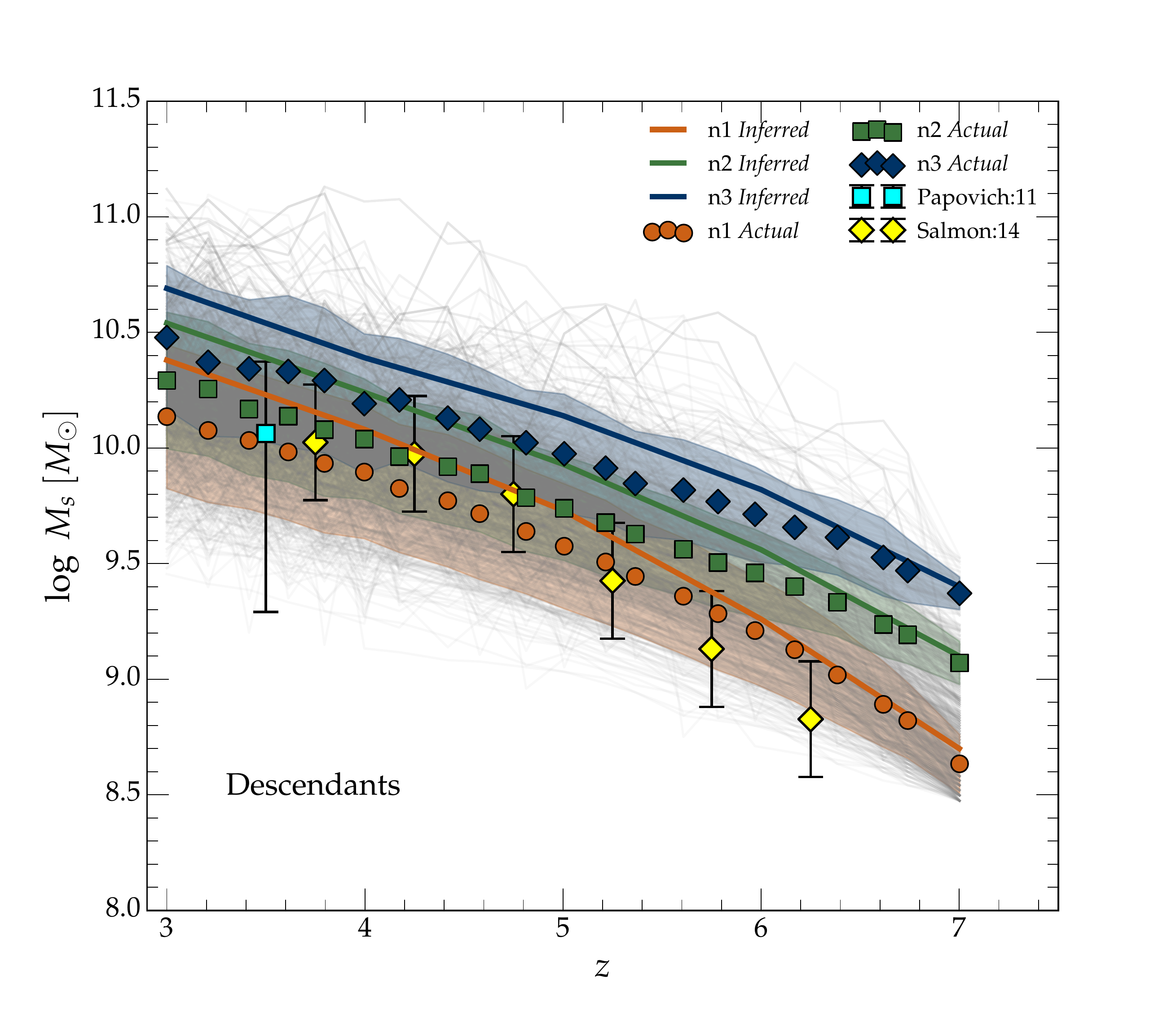}
\caption{{\it Actual} mass growth for descendants, with individual galaxies shown as the gray lines, and medians in redshift bins shown by the orange/green/blue circles/squares/triangles, compared to the {\it inferred} mass growth by tracking a constant number density from the CSMF (solid orange/green/blue lines).  The shading represents the standard deviation of the stellar mass at a given redshift of the individual simulated galaxies.  Observed results from \citet{Papovich.etal:11} (cyan square) and \citet{Salmon:15} (yellow diamonds) are show for comparison.  These observations were obtained from the same number density as our n1 bin.   While the inferred mass growth does appear to track the average mass growth of the individual simulated galaxies, there is both a bias of 0.2 dex and a substantial scatter of $\pm 0.30$ dex by $z =$ 3.
}
\label{fig:m_evo_d}
\end{center}
\end{figure*}

\begin{figure}
\centerline{\includegraphics[width=1.1\columnwidth,angle=0] {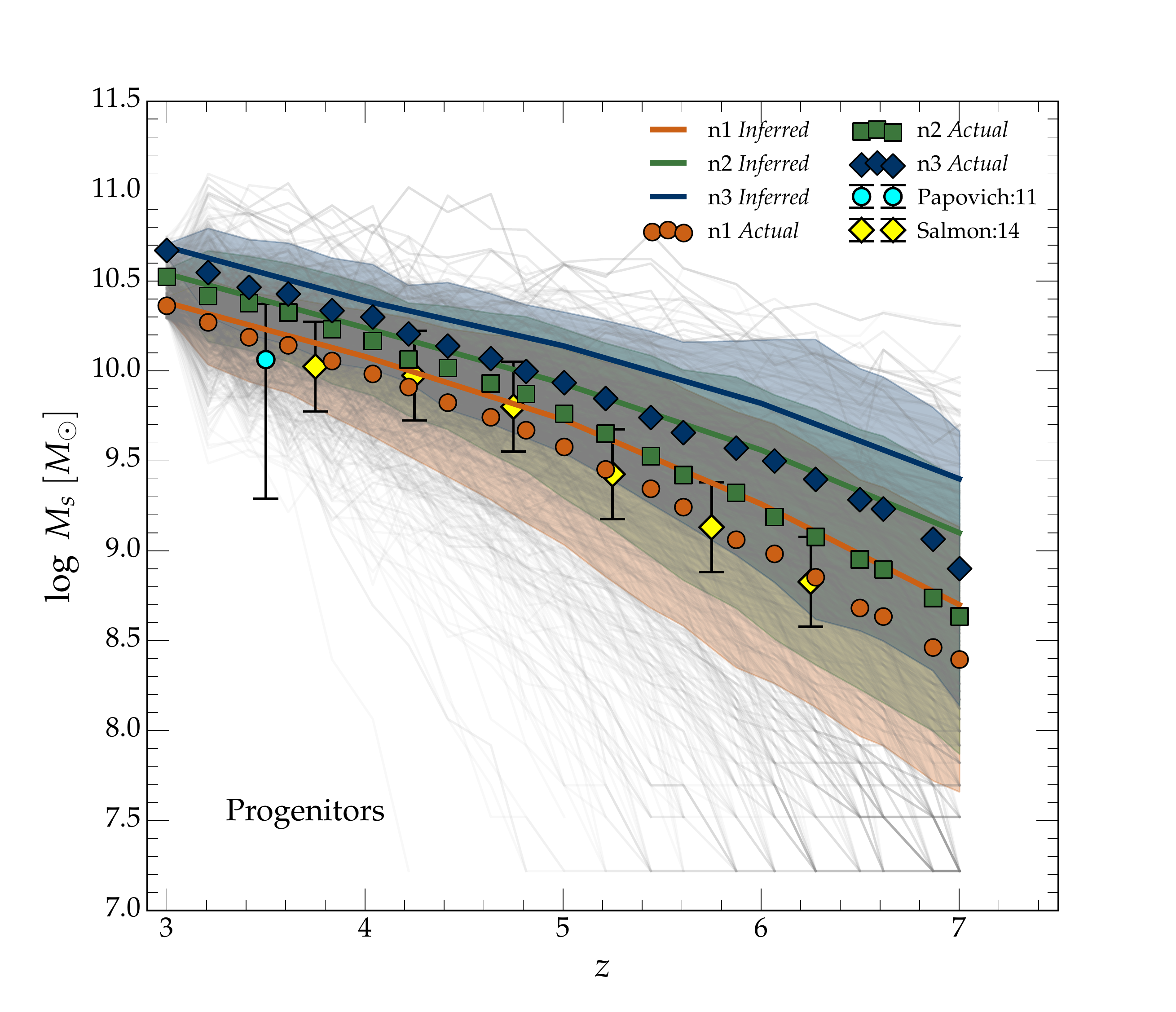}}
\caption{{\it Actual} mass growth for progenitors, compared to the {\it inferred} mass growth from the CSMF.  All symbols, lines and shading represent similar quantities as in Figure~\ref{fig:m_evo_d}, though here for progenitors of $z =$ 3 galaxies.  While again this technique appears to work on average, both the bias and the scatter of the constant number density tracking technique are larger when tracking progenitors than when tracking decendants.}
\label{fig:m_evo_p}
\end{figure}

While the algorithm for tracking galaxies in the simulation is simple, in practice this is a non-trivial exercise.  All group-finders introduce some amount of uncertainty when associating particles with a galaxy, especially during merger events.  Particles which were included in one snapshot can be excluded in the subsequent one, primarily those which are located more near the edges of the identified system. This uncertainty manifests itself as ``noise'' in the mass growth of each galaxy (see jagged gray lines in Figure.~\ref{fig:m_evo_d}).  While this ``noise'' is aesthetically unpleasing, it does not affect the overall median growth trend.

\subsection{Actual versus inferred mass growth}

Throughout this work we will refer to the {\it actual} mass growth and the {\it inferred} mass growth of the studied galaxy population.  The {\it inferred} growth refers to the mass assembly of the ensemble galaxy population as derived from the CSMF by taking the mass associated with a constant number density cut at each redshift (e.g. the orange line intersection with CSMF in Figure~\ref{fig:cmf}).  The {\it actual} mass growth is the mass assembly unique to each individual galaxy (gray lines in Figure.~\ref{fig:m_evo_d}) determined at each snapshot as described in Sec.~\ref{sec:tracking}.  In each case the {\it actual} mass growth will be represented by the median of the distribution with the $1\sigma$ scatter represented by shading (e.g. orange circles, orange shade Figure.~\ref{fig:m_evo_d}).


\section{Results}
\label{sec:results}
\subsection{Descendants}
\begin{figure*}
\begin{center}
\includegraphics[width=1.00\columnwidth,angle=0] {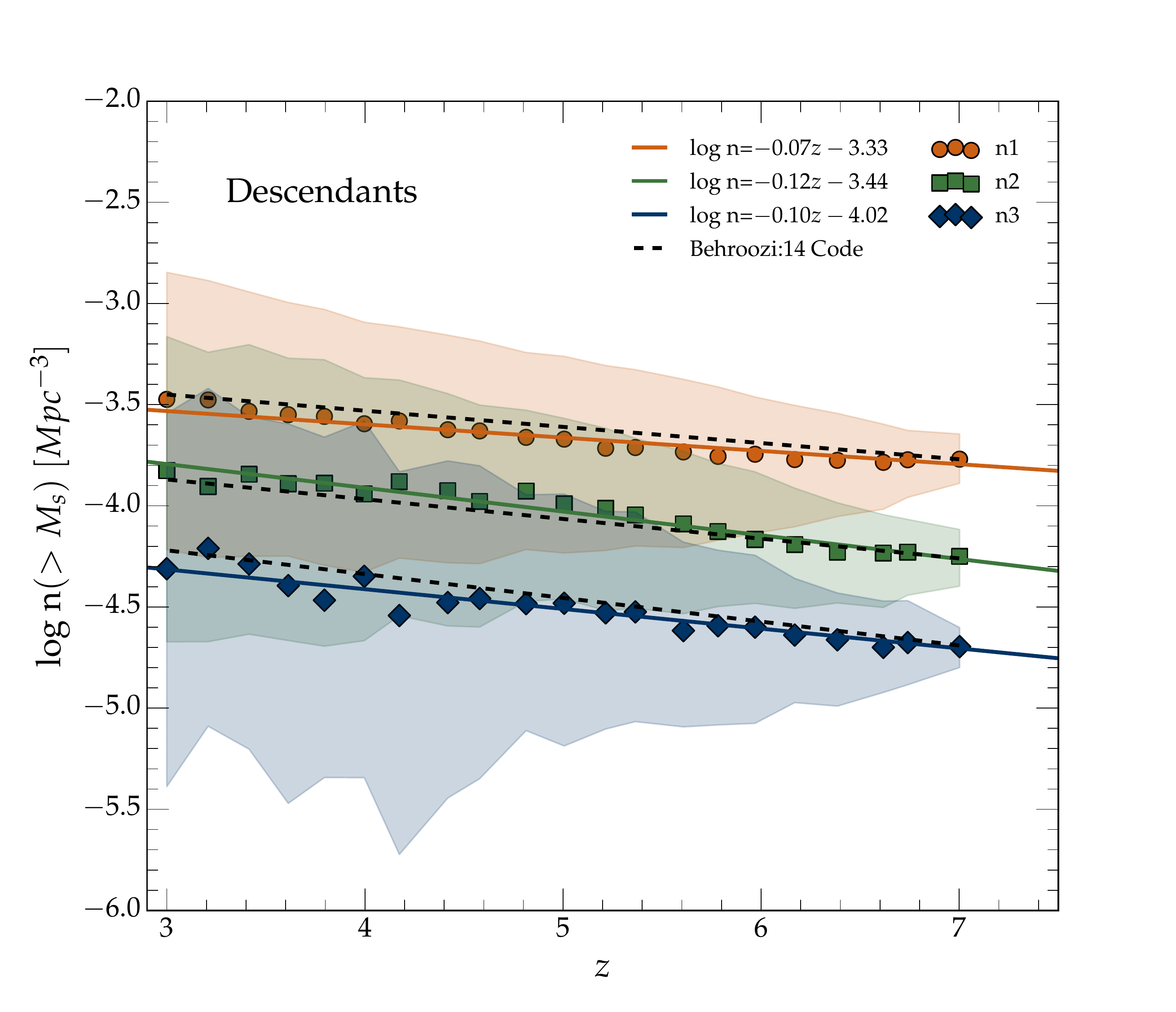}
\includegraphics[width=1.00\columnwidth,angle=0] {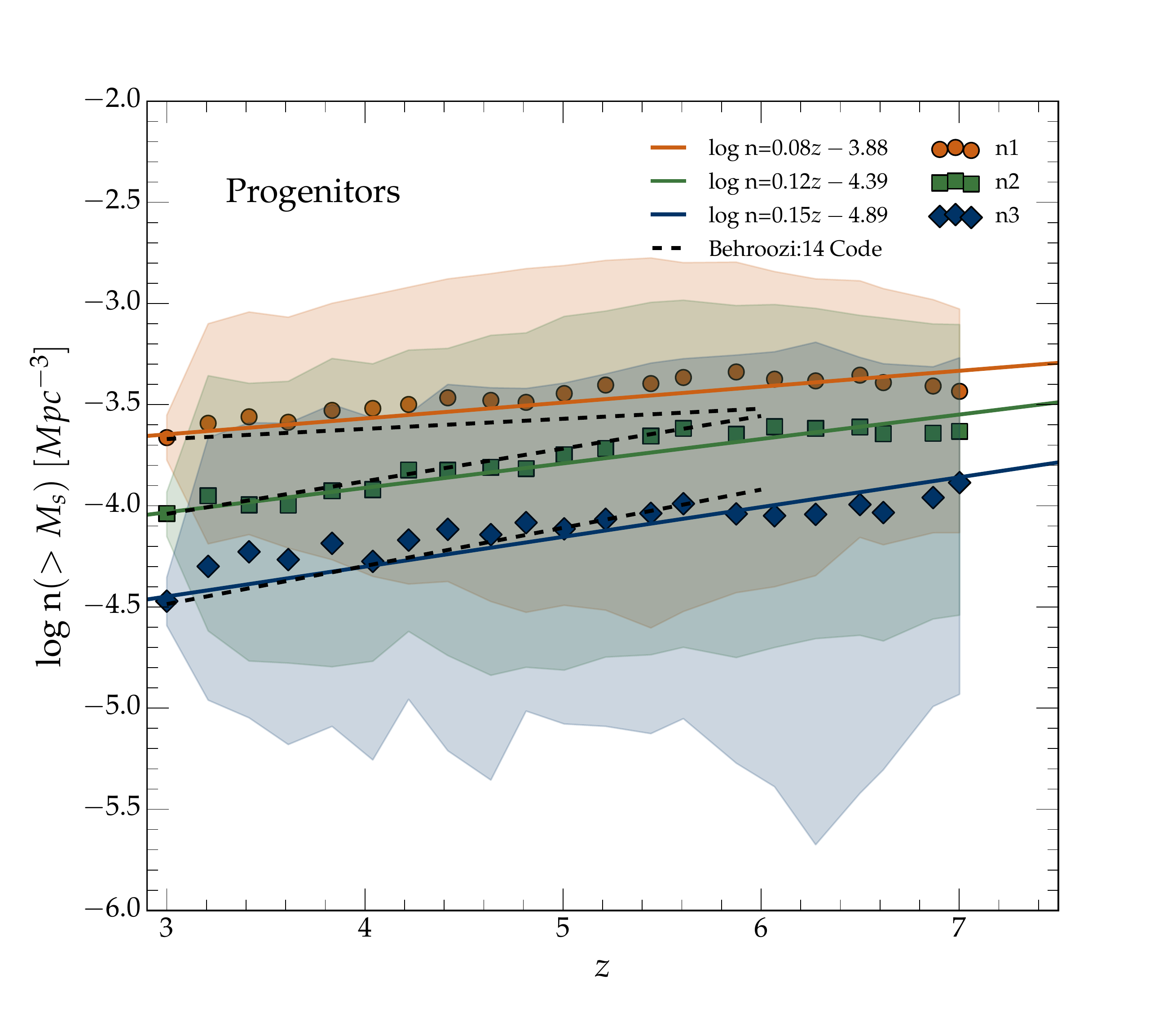}
\caption{Redshift evolution of the number density for each population (n1, n2, and n3) for both descendants (left) and progenitors (right), with the shaded regions denoting the $1\sigma$ scatter.  For the progenitor plot (right), galaxies that form after $z$=7 are dropped at their formation redshift as we approach higher redshifts and therefore do not contribute to the scatter prior to their formation ( i.e., the total number of galaxies is reduced, and the average mass does not include objects with zero stellar mass; see Section~\ref{subsec:scatter} for more details).  This makes the scatter information less reliable at higher redshifts for progenitors.  We provide linear least square fits (solid lines) to the median evolution for each population (solid orange, green, blue lines) which can be directly compared with the results from \citet{Behroozi:13}, shown as the dashed lines, showing excellent agreement, in spite of the two very different techniques between our two studies.}
\label{fig:dens_evo}
\end{center}
\end{figure*}

In Figure~\ref{fig:m_evo_d} we compare the inferred mass growth to the median of the actual mass growth for each galaxy in the n1, n2, and n3 bins.  This comparison illustrates that the {\it inferred} growth tracks the {\it actual} mass growth on average, though there appears a systematic bias where the inferred mass growth is $\sim0.20$ dex higher (within a factor of $<2$ in mass) than the actual mass growth.  
This demonstrates that, on average, using constant cumulative number density to track the descendants of a population of galaxies through cosmic time can be an effective method of measuring stellar mass growth.  This result is consistent with studies using the same techniques at lower redshifts \citep{vandokkum:10,leja:13} and at high-$z$ \citep{Papovich.etal:11,Behroozi:13}.   We caution however that this is an {\it on average} result which has a large scatter ($\sim \pm0.30$ dex) which we will discuss in more detail in Section~\ref{subsec:scatter} and a systematic offset which suggest that evolution of the population must be considered (see Section~\ref{sec:evo}).

In Figure~\ref{fig:m_evo_d} we also compare to stellar mass growth estimates obtained through observations by \citet{Salmon:15} and \citet{Papovich.etal:11}.  This comparison demonstrates that our simulations, on average, are reproducing the observed growth of galaxies selected from a bin similar to our n1.   As \citet{Salmon:15} results were obtained using the evolving cumulative number density technique, it also serves as a good check that our simulations are reproducing the observed galaxy properties and evolution well.

\subsection{Progenitors}
\label{sec:progen}

We also use our simulations to test the effectiveness of the constant cumulative number density technique to track the \emph{progenitors} of a population of galaxies selected at $z$=3.  For the purposes of this study we are defining a galaxy progenitor to be it's most massive progenitor.  We acknowledge that in a hierarchical galaxy formation scenario each galaxy is the aggregate of many progenitors of varied masses.  However, from an observational perspective, it is impossible to link multiple progenitors to a single descendant.  Therefore we feel that our definition more closely approximates what can be observed.  Figure~\ref{fig:m_evo_p} shows the mass assembly history of $z =$ 3 galaxy populations in the n1, n2, and n3 bins.  As with the descendants, the constant cumulative number density technique, {\it on average}, reproduces the mass assembly history of the population of galaxies when compared to the {\it inferred} assembly, although with a larger bias of $0.30$ dex (factor $\sim2$ in mass).  

However, when using this technique to track progenitors the scatter in the individual galaxy history increases substantially to $\sim \pm0.70$ dex at $z$=7 (Figure~\ref{fig:m_evo_p}), and is much higher than was seen when tracking descendants ($\sim 0.30$ dex).  Further, we find that $\sim30\%$ of the galaxies in the original number density bin at $z$=3 do not exist at $z$=7, forming at various points between $z$=3 and $z$=7.  This population of ``late-forming'' galaxies substantially increases the scatter in the {\it actual} assembly history and thus increases the offset between the {\it actual} and {\it inferred} mass growth.   It should be noted that we drop the late-forming galaxies from the calculation of average mass growth prior to their formation redshift to avoid including objects with zero stellar mass.  We include more discussion regarding the impact of this ``late-forming'' galaxy population in Section~\ref{subsec:late}.

\subsection{Evolving number density}
\label{sec:evo}

The systematic offset between the {\it actual} and {\it inferred} mass growth suggests that the number density of the original population is evolving with redshift.  This evolution can be seen in Figure~\ref{fig:dens_evo} which shows the evolution of the number density as a function of redshift for both descendants (left) and progenitors (right), for each of our initial number density bins.  The number density for each individual galaxy is obtained at each snapshot by relating its stellar mass to a number density via the CSMF.  All populations show a trend of increasing number density as a function of cosmic time (with decreasing redshift for descendants, and with increasing redshift for progenitors).  This demonstrates that the average {\it actual} mass growth of galaxies from each number density bin are growing slower than would be inferred from the CSMF (see offset between {\it inferred} and {\it actual} stellar mass growth in Figures~\ref{fig:m_evo_d} \& \ref{fig:m_evo_p}).

This trend is highly consistent with findings from  \citet{Behroozi:13} who utilized the Bolshoi \citep{Klypin:11} dark matter simulations and  abundance matching (see Section~\ref{subsec:abund} for further details).  In Figure~\ref{fig:dens_evo} we compare our linear best fit (solid lines) to the results from the public code provided by \citet{Behroozi:13}; dashed lines.  Our work shows excellent agreement (within $\sim 0.10$ dex) with the evolution generated by the \citet{Behroozi:13} code over the studied redshift range.  

Thus, in order to track galaxies more accurately through cosmic time one must utilize an {\it evolving} cumulative number density. By taking the average of the least square fit to the evolutionary trend for each population (see least square fits in Figure~\ref{fig:dens_evo}) of descendants of $z$=7 galaxies ($10^{8.48}\leq M_s/\Msun \leq 10^{9.55}$) we find,
\begin{equation}
\label{eq:one}
\log n_f = \log n_i-0.10 \Delta z
\end{equation}
Similarly the number density of $z$=3 ($10^{10.39}\leq M_s/\Msun \leq 10^{10.75}$) progenitors can be traced utilizing,
\begin{equation}
\label{eq:two}
\log n_f = \log n_i+0.12 \Delta z
\end{equation}
We caution that these relations are only valid in the mass range and redshift range studied here ($3\leq z \leq7$).  As pointed out in \citet{Behroozi:13}, the evolution changes at $z\leq3$, therefore equations~\ref{eq:one} \& \ref{eq:two} should not be extrapolated to lower redshifts.  Note that the mass ranges referenced above are derived from the overall number density range (n1, n2, n3) at either $z=$7 for descendants or $z=$3 for progenitors (see Table~\ref{tbl:data}).

\subsection{Completeness}
\label{subsec:comp}

\begin{figure}
\centerline{\includegraphics[width=1.0\columnwidth,angle=0] {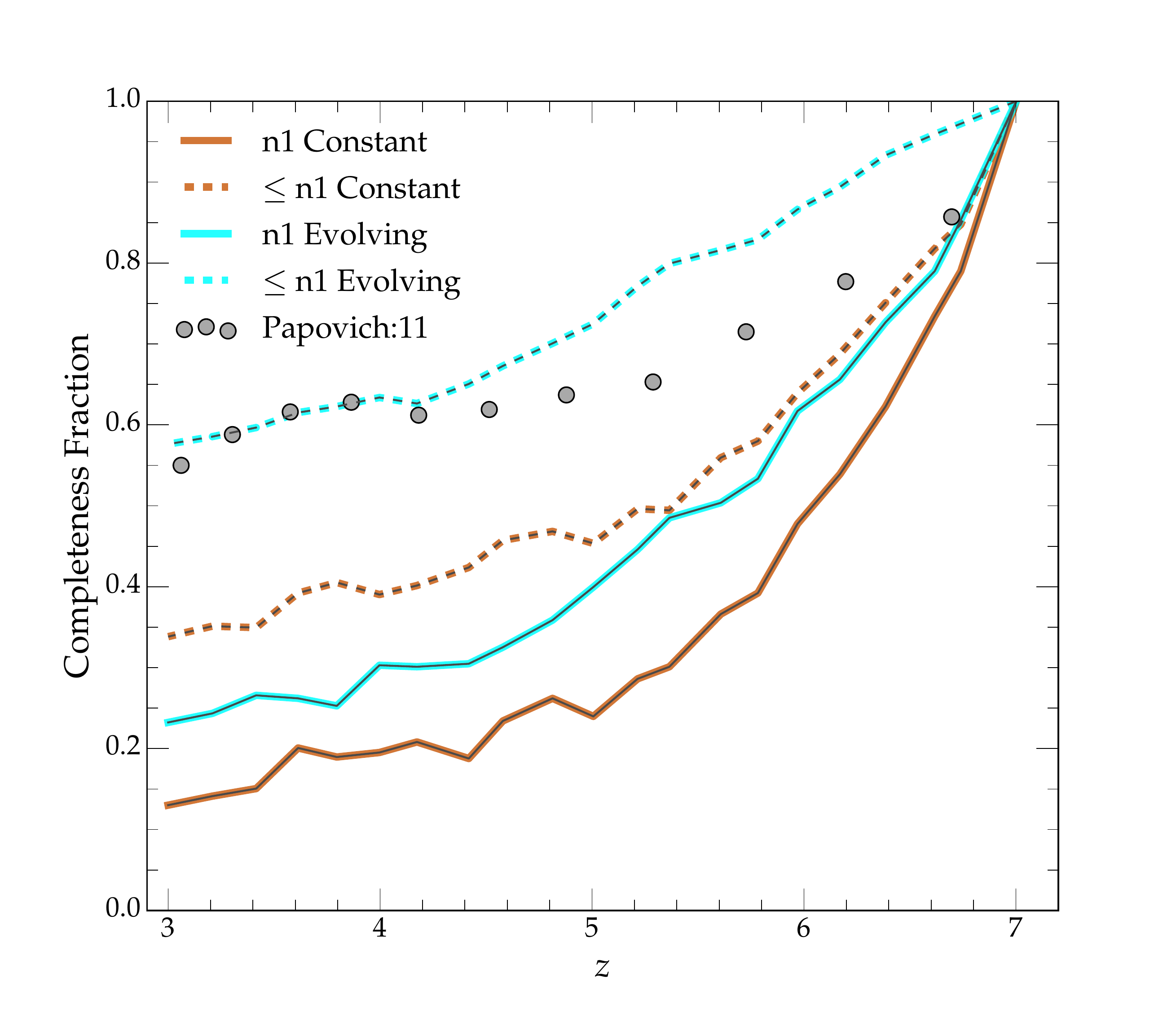}}
\caption{Completeness fraction, defined to be the fraction of tracked galaxies at each redshift which fall within the desired number density bin, keeping the bin width constant.  The solid orange line represents the completeness fraction for the n1 bin.  We also compare to \citet{Papovich.etal:11} (gray circles), who define the completeness as the fraction of galaxies with  a number density less than or equal to the value of the selection bin.  Our direct comparison to the published data is the dashed orange line.  The cyan solid and dashed lines represent the completeness fraction after applying the evolving cumulative number density, both using our definition of completeness, and that of Papovich et al.  We find that even after the correction for the evolving number density, only $23\%$ of the $z=$7 population is recovered at $z=$3.}
\label{fig:comp}
\end{figure}

One method which has been used to quantify how well the constant cumulative number density method is tracking the evolution of the galaxy population is to examine the completeness fraction at each redshift, keeping the bin width constant.  Here we define the completeness fraction to be the fraction of tracked simulated galaxies which fall within the desired number density bin at a given redshift.  In Figure~\ref{fig:comp} we present the results of this study, where we find that only $\sim 13\%$ of $z$=7 descendants fall within the original number density bins at $z=3$.  For clarity, we only show results for the n1 bin, however, the results for the n2 and n3 bins are quantitatively similar.  We do not present the completeness fraction for progenitors for reasons which we discuss in Section~\ref{subsec:scatter}.

We also compare our results to those found in \citet{Papovich.etal:11}.  To make this comparison we need to modify our definition of completeness to match that used in Papovich et al.  They define completeness as the fraction of galaxies with a number density of at most the value of the selection bin ($n(>M_s)$).  Our direct comparison to this definition can be seen by the dashed orange line.  This definition results in a completeness fraction of $\sim 35\%$, much lower than the \citet{Papovich.etal:11} result of $\sim 55\%$.  To obtain a similar result to the \citet{Papovich.etal:11} study we would need to expand our target mass bin at $z$=3 to $\pm0.29$ dex.

Repeating both of the completeness calculations above, now applying the evolving cumulative number density leads to a completeness fraction of $\sim23\%$  as defined in this work and $\sim58\%$ for the \citet{Papovich.etal:11} definition.  This makes our completeness fraction nearly identical to the \citet{Papovich.etal:11} result using their less restrictive definition.  However, we see a modest increase of only $10\%$ at $z$=3 after applying the evolving cumulative number density when using our completeness definition.   This discrepancy is likely due to the larger scatter between halo mass, galaxy mass and luminosity found in our simulations.  The study by \citet{Papovich.etal:11} did not include any of these sources of scatter in their estimates.

The results from the completeness fraction study are somewhat misleading as the width of the $z$=3 mass bin associated with the selected number density is smaller than the original $z$=7 mass bin by a factor of $\sim 2$ (see Table~\ref{tbl:data}).  The reduced mass bin width is the result of the steepening of the CSMF over the redshift range studied, and it is the primary reason why we only see a modest completeness fraction increase of $10\%$ at $z$=3 when applying the evolving cumulative number density.  If the mass bin width is fixed at the $z=$7 value of $\sim\pm 0.20$ dex (see Table~\ref{tbl:data}), the completeness fraction at $z=$3 rises to $\sim55\%$.

\subsection{Scatter}
\label{subsec:scatter}
\begin{figure*}
\begin{center}
\includegraphics[width=1.5\columnwidth,angle=0] {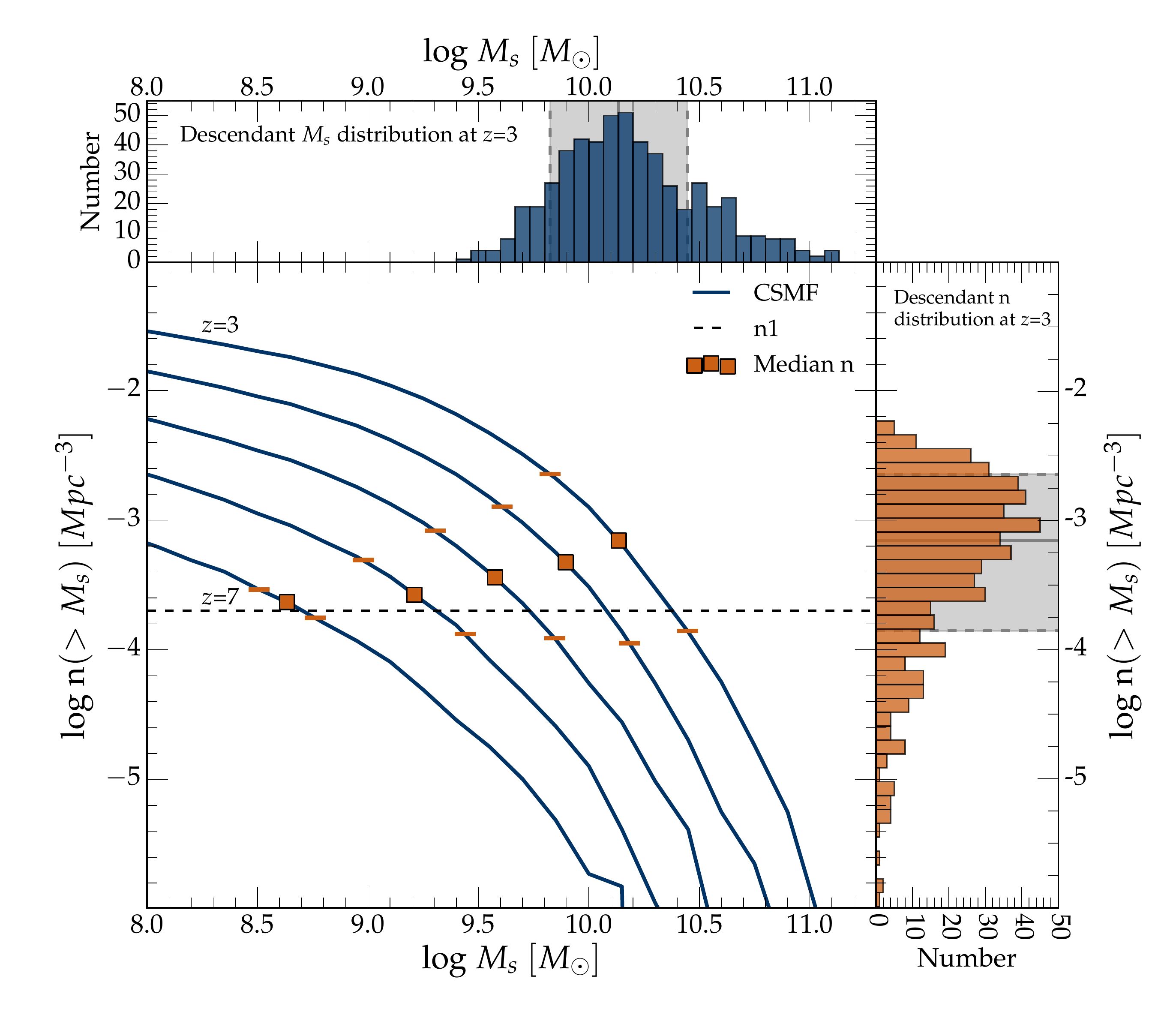}
\caption{CSMF for $z$=7--3 simulated galaxies (blue solid lines, left to right) shown with the evolution of the median number density of the descendants of $z =$ 7 galaxes in the n1 bin (orange squares).  The orange horizontal hash lines represent the 1$\sigma$ scatter around the median at each redshift.  The histograms illustrate the distribution of the simulated descendants in both number density (orange) and stellar mass (blue) at the terminal redshift of $z =$ 3.  While the distribution is wide there does exist a clear peak in both number density and mass. }
\label{fig:dscatter}
\end{center}
\end{figure*}

\begin{figure}
\centerline{\includegraphics[width=1.1\columnwidth,angle=0] {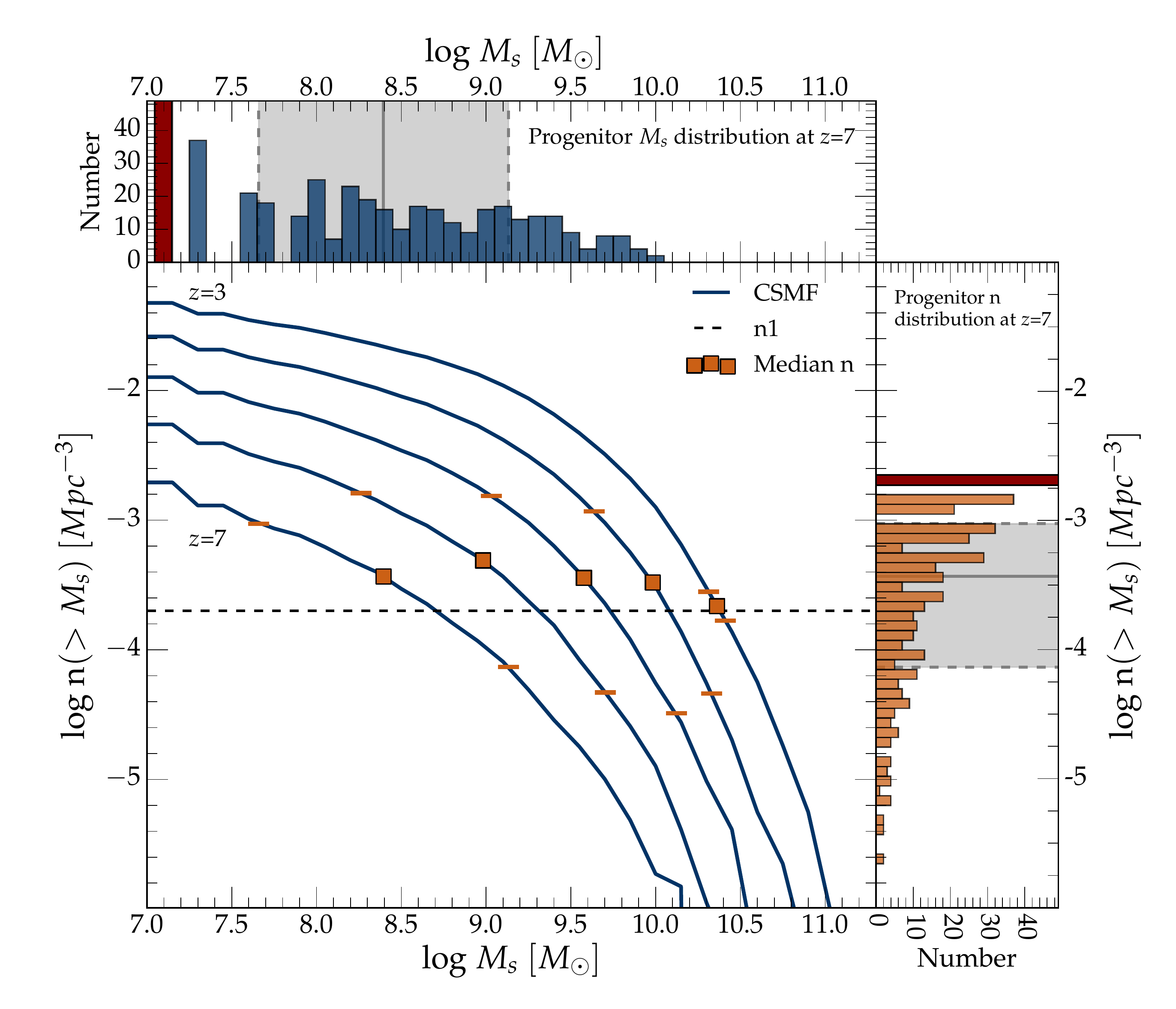}}
\caption{The same lines and symbols as Figure~\ref{fig:dscatter}, only here for progenitors of $z =$ 3 galaxies tracked to $z =$ 7.  The red histogram bar in the progenitor panel represents the population ``late-forming'' galaxies found in the progenitor study.  Progenitors of $z=$3 galaxies have a much wider distribution in both mass and number density that seen from descendants of $z=$7 (see Figure~\ref{fig:dscatter}).}
\label{fig:pscatter}

\end{figure}

As demonstrated by the low completeness fraction, the majority of galaxies selected at $z$=7 are not falling within either the evolved or constant number density target bin at $z$=3. It is therefore useful to constrain where the majority of these descendants fall on the CSMF.  Figures~\ref{fig:dscatter} and \ref{fig:pscatter} shows the median number density evolution of simulated galaxies in our n1 bin plotted with the CSMF at each redshift for descendants and progenitors, respectively.  These figures synthesize much of the data provided in this work as it also includes information regarding the mass and number density distribution (blue/orange histograms) for both descendants and progenitors. 

For the n1 bin descendants, we find that $\sim 11\%$ of galaxies fall in the n1 bin at $z$=3, as discussed in Section~\ref{subsec:comp}, and the remaining galaxies are distributed either at higher number densities ($\sim 67\%$), or at lower number densities ($\sim 22\%$) than the original number density bin.  The overall scatter in stellar mass at $z$=3 is $\pm0.32$ dex, which corresponds to a number density scatter which has a total width of $\sim 1.2$ dex, given the $z =$ 3 CSMF.  For progenitors of the n1 bin, we find that $\sim 13\%$ fall within the target $z =$ 7 number density bin with $\sim67\%$ at higher number densities, and $\sim21\%$ at lower number densities.  This distribution leads to a stellar mass scatter of $\pm0.75$ dex at $z$=3, and a number density scatter with total width of $\sim1.1$ dex.  It should be noted that the ``late-forming'' galaxies discussed in Section~\ref{sec:progen} are included in the fraction of galaxies which fall above the n1 bin.  However, they are removed from the sample at their formation redshift.  Therefore, they do not contribute to the median scatter (gray shaded area in Figure~\ref{fig:pscatter}) at $z=$3.  The n2 and n3 density bins have quantitatively similar results.

 It is interesting to consider the growth histories of galaxies which started in the selected number density bin at $z$=7, and end up outside the bin at $z$=3.  We performed a test to determine the average growth rates of these populations of galaxies, calculating the average stellar mass at each time step of galaxies outside the bin (separately for those with both higher and lower number densities). We find that these \emph{average} growth histories show that these galaxies initially grow too quickly or too slowly (moving to lower or higher number densities, respectively), jumping out of the selected number density bin, but after that, they grow at a relatively constant rate, similar to those galaxies inside the selected bin.  Therefore, there is no systematic difference in the growth histories of these galaxies; rather, after they initially scatter out of the number density bin, they grow fairly similarly to the tracked population.

The fact that both the descendants and the progenitors have similar fractions of galaxies located in, above and below the original bin is misleading.  As mentioned in Section~\ref{subsec:comp}, the slope of the CSMF is steepening with decreasing redshift, resulting in a narrower number density bin at $z$=3 and a lower descendant completion fraction.  This must also be considered when looking at progenitors as the bin is widening as we approach $z$=7, thus making it more likely that a galaxy will fall within the selection bin (i.e. higher completion fraction).  

As demonstrated by the mass and number density histograms included in Figure~\ref{fig:pscatter}, even using an evolving cumulative number density selection results in a wide scatter in both number density and mass for progenitors.  We conclude that using an evolving cumulative number density will allow for the tracking of {\it average} stellar mass for progenitors, but that this method is an insufficient tracer of individual galaxy descendant/progenitor evolution in our simulations, as nearly any bin on the $z$=3 CSMF can be populated with galaxies that originated from nearly any point on the $z$=7 CSMF.  We show further evidence of this scatter and its impact in Section~\ref{subsec:disc_scatter}.

\section{Discussion}
\label{sec:disc}
\subsection{Scatter}
\label{subsec:disc_scatter}

\begin{figure}
\centerline{\includegraphics[width=1.00\columnwidth,angle=0] {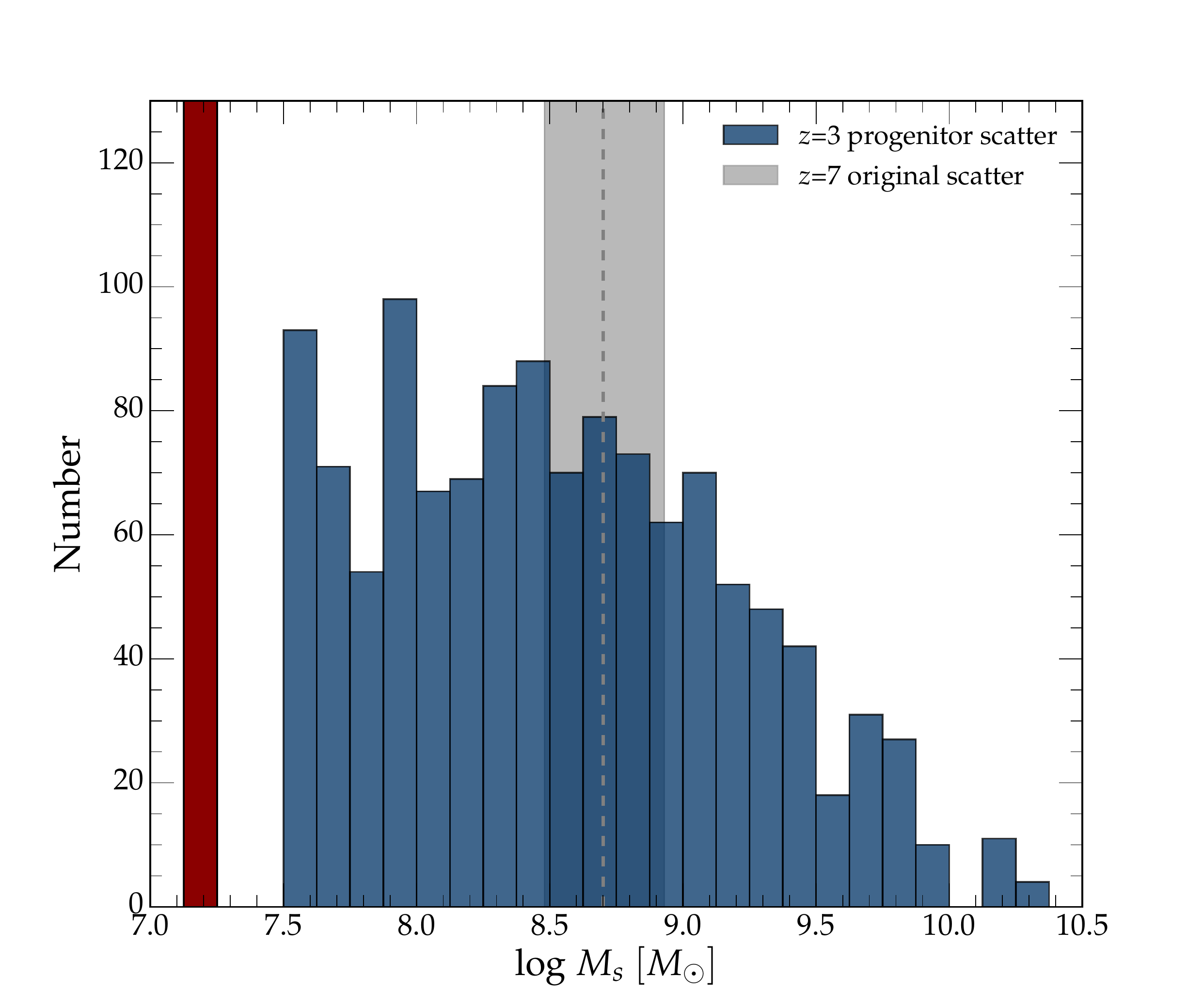}}
\caption{The mass distribution of $z =$ 7 simulated galaxies, which were tracked as progenitors of $z =$ 3 galaxies, which were themselves tracked as descendants of $z =$ 7 galaxies in the n1 number density bin (effectively a ``round-trip'' tracking exercise).  Here we show the stellar mass distribution for \emph{all} galaxies selected in a mass bin at $z =$ 3 which contained 50\% of the originally-tracked $z =$ 7 descendants (see Section~\ref{subsec:comp} for details).  The gray shading indicates the original n1 bin width and the red bar represents the late-forming population found during progenitor tracking (see Section~\ref{sec:progen}).  This ``round trip'' tracking exercise demonstrates that a given point on the $z$=3 CSMF can be populated by galaxies which have origins at nearly any point on the $z$=7 CSMF.}
\label{fig:50}
\end{figure}

To further explore the implications of the large scatter in descendants, an exercise was conducted to determine the mass bin width required at $z$=3 to obtain a completeness fraction, as defined in this work, of $50\%$.  It was found that an increase of a factor of $\sim 3$ from $\pm0.10$ dex in stellar mass to $\pm0.29$ dex in stellar mass at $z$=3 is required to recover $50\%$ of the original $z$=7 selected population (see Table~\ref{tbl:data}).  While a mass bin with a width of $\pm0.29$ dex may not seem extreme, as it is comparable to the observed mass error estimates at $z=3$, it will contain thousands of galaxies that were not part of the original $z$=7 density bin, which may have very different mass assembly histories.  This can be see in Figure~\ref{fig:50} where the stellar mass distribution for a sample of $2175$ galaxies were selected from the aforementioned $50\%$ bin.  The progenitors of this sample were then tracked back to $z$=7 for comparison to the original $z$=7 population.  The blue histogram represents the $z =$ 7 stellar mass distribution of the $2175$ $z$=3 progenitors, and the gray shaded region indicates the mass width of the original n1 number density bin.  This exercise effectively demonstrates that even a moderate widening of the target mass bin will lead to substantial contamination from a diverse population of galaxies.  This exercise also shows that a large scatter makes the application of either the constant cumulative number density or evolving cumulative number density technique limited to average population masses only.  The diverse mass assembly histories of individual galaxies can make associating other properties such as age, metallicity, star-formation rate, etc., untenable.

\subsection{Late-forming galaxies}
\label{subsec:late}
The $\sim30\%$ of $z$=3 progenitor galaxies found in our simulation to form after $z$=7 have a significant impact on the results of the progenitor portion of this study.  In particular, the scatter of the stellar mass and number density evolution is increased by a factor of $\sim2$.  The high fraction, and significant impact of the ``late-forming'' population dictate that further discussion is warranted.  Specifically, there is the possibility that these galaxies are artifacts of our simulations due to resolution limitations.  The large volume of our simulation box is chosen so that we can best represent the dynamic range of observed high redshift galaxies.  The consequence of this design is that we are not resolving low-mass halos with $M_h \lesssim 10^9 \Msun$ at $z>$7.  Therefore, the possibility exists that a theoretical simulation box with unlimited mass resolution would have resolved these ``late-forming'' galaxies at an earlier redshift.  

In a $\Lambda$CDM Universe with a hierarchical structure formation scenario, it is possible that these galaxies would have then been merged into a higher mass galaxy by $z$=7, thus reducing the scatter.  It is also possible that no mergers took place which would leave the galaxy in a lower mass bin (i.e. much higher on the CSMF). In this scenario the scatter would remain largely unchanged.  To properly study this issue, a resolution test would be needed, which requires a simulation with a substantially higher particle count.   Unfortunately, such a hydrodynamic simulation is beyond the scope and resource of this study. 

There is also the possibility that the ``late-forming'' galaxies are very ``real'', and representative of early galaxy formation.  A simple calculation, exploring the growth rate required for a galaxy at $z$=5 with $M_s\approx10^{7.1}\Msun$ to reach the target stellar mass of $M_s\approx10^{10.4}\Msun$ by $z$=3, demonstrates that such a galaxy would require a star formation rate (SFR) of $\sim25\ \Msun\ yr^{-1}$ over the $\sim 1$ Gyr.  This SFR is reasonable, given that a typical observed $L_*$ galaxy has an SFR$\approx50\ \Msun\ yr^{-1}$ at $z$=5 \citet{Finkelstein:14} .  This is without consideration for mergers, which could increase the stellar mass much more rapidly.  Therefore, there is no physical motivation for us to definitively conclude that the ``late-forming'' galaxies are mere resolution artifacts. 

We investigated the photometric properties of the putative ``late-forming'' galaxies to determine if it would be possible to detect $z$=3 observational signatures.  More specifically, we examined the D4000 spectral indicator, which measures the relative strength of filters placed around the $4000\text{\AA}$ spectral features associated with Calcium H and K lines \citep{Bruzual:83}.  The strength of this break is correlated with age, as younger, high SFR galaxies will produce an increase of UV photons at wavelengths less than $4000\text{\AA}$.  The spectra of these galaxies will also contain less observable metal-line absorption, which will result in a decreased D4000 ratio.  Utilizing the BC03 \citep{BC03} population synthesis models to generate spectra for each of our simulated galaxies, we found that there is no significant difference in the D4000 break for ``late-forming'' galaxies as compared to the remaining studied population.  It is likely that this is the result of a combination of merger events and rapid star formation that leads to the effective mixing of older stellar populations with the newly formed populations making the $4000\text{\AA}$ an unreliable age indicator for our galaxy population.

\subsection{On the origin of the evolving number density}
\label{subsec:evo}
When interpreting the results of this work, and other similar studies, it is critical to remember that number density is not a physical property of a galaxy (i.e. unlike stellar mass, SFR, metallicity, etc.).  The number density of any individual galaxy is dependent directly on the distribution function from which it is taken, in this case the cumulative stellar mass function.  As seen in Figure~\ref{fig:cmf}, the CSMF is evolving with decreasing redshift towards more galaxies at higher masses, but also toward a stronger high-mass exponential cut-off.  There are two primary forces driving this evolution of the CSMF over the studied redshift range:  First, the stellar masses of existing galaxies are growing, which evolves the CSMF to toward higher masses.  Second, new galaxies are forming and being included which, in a ``bottom-up'' hierarchical structure formation universe, will place them at higher number densities, effectively acting to steepen the exponential cutoff of the CSMF.  We caution that this steepening is not related to the faint-end slope of the stellar mass function as our dynamic range does not include faint-end mass galaxies.

The steepening of the CSMF cutoff directly results in the evolution of the cumulative number density in our simulations.  This can be clearly demonstrated by examining the rise in cumulative number density for a given stellar mass bin as the exponential cutoff becomes more pronounced (i.e. towards vertical). Therefore, a population of galaxies can have smoothly rising mass assembly histories, as seen in Figure~\ref{fig:m_evo_d}, with no break in rank order, and still exhibit an evolving number density.  This result is contradictory to the fundamental assumptions that are made in the constant cumulative number density technique, which has been shown to work at lower redshift  \citep[$0\leq z\leq3$;][]{vandokkum:10,leja:13}.  We do not however feel that our results invalidate the lower redshift results as the processes driving galaxy evolution are different during that more recent epoch (i.e. driven by stellar mass growth and not new galaxy formation, with an increased impact of feedback on massive galaxies). 

Based on the above evolution argument, one would expect that the progenitor number density evolution would then trend towards lower number densities at higher redshifts.  We do not see this trend.  In fact we find the opposite, with progenitors evolving to higher number densities (Figure~\ref{fig:pscatter}, bottom).  This can be explained by the impact of the large ``late-forming'' galaxy population acting to lower the median stellar mass growth (Figure~\ref{fig:m_evo_p}), thus raising the cumulative number density.

It must be noted that the steepening of the CSMF in our simulations could also be the result of our mass resolution limitations as discussed in Section~\ref{subsec:late}.  Low mass galaxies which are unresolved would fall at higher number densities, and could therefore steepen our $z=7$ CSMF, potentially eliminating the evolution seen in this work.  Cosmological volume hydrodynamic simulations with sufficient dynamic range to test this effect are beyond the scope of this work based on the computational resources required.

\subsection{Implications of scatter for abundance matching}
\label{subsec:abund}
The agreement with the \citet{Behroozi:13} results is rather remarkable given the differing approaches taken to arrive at these conclusions.  In the \citet{Behroozi:13} work an abundance matching technique is employed.  This technique converts the cumulative number density corresponding to a galaxy's stellar mass at a given initial redshift to a dark matter halo mass with the same cumulative number density from the dark matter-only Bolshoi simulation halo catalog.  The progenitors/descendants of these halos are then tracked within simulation using a similar `most-massive' algorithm \citep{Behroozi:13c}.  At the terminal redshift the mass of each halo is then converted back to a cumulative number density using the dark matter halo cumulative mass function \citep{Behroozi:13b}.

At the core of the abundance matching technique is the fundamental assumption that halo mass accretion, whether it be through mergers or via the cosmic web, directly traces the galaxy stellar mass growth hosted within.  This is a reasonable assumption given that we know that the baryonic matter in the Universe directly traces the underlying dark matter distribution \citep{Blanton:99}.  It does not however account for the complicated physics of how this gas accretes onto galaxies and ultimately is converted into stars.  This absence of fundamental physics is the primary motivation for this work.

The utilization of cosmological SPH simulations is the next logical step in testing these galaxy-tracking techniques, as they include both the underlying dark matter structure and the complicated hydrodynamic physics associated with the gas within (i.e. heating, cooling, chemistry).  These simulations are also able to convert gas into stars via physically motivated sub-grid physics as well as account for the impact of star formation (i.e. supernova and stellar winds) on further star formation.  This allows for galaxies to form in-situ with their respective dark matter halos, while computing the gas infall rate from cosmic filaments in a cosmological setting correctly..  Therefore we are able to directly track galaxy stellar mass growth without the need to make any assumptions regarding the relationship between the galaxy and the halo it resides within.  The drawback to utilizing these types of high resolution SPH simulations is that they are computationally very expensive.  This limits the dynamic mass range and redshift range we are able to study when compared to dark matter only simulations (i.e. Bolshoi, Millennium), which resolve lower mass structures.  To probe this topic further, a combination of theoretical techniques is likely warranted.

\section{Summary}
\label{sec:sum}

Using the cosmological smoothed particle hydrodynamical code {\small GADGET-3} we make a realistic assessment of the validity of the technique of using constant cumulative number density as a tracer of galaxy evolution at $3\leq z \leq7$.  Our major conclusions are as follows:

\begin{itemize}
\item We find that, when utilizing the constant cumulative number density technique to track galaxies at $3\leq z \leq7$, an average stellar mass can be determined to within a factor of $<2$ for descendants and $\sim2$ for progenitors (Figure~\ref{fig:m_evo_d} \& \ref{fig:m_evo_p}).

\item The systematic offset between the {\it inferred} mass growth and the {\it actual} mass growth suggest that the number density of both progenitors and descendants is evolving with redshift.  This evolving cumulative number density is in very good agreement with \citet{Behroozi:13} who find a similar evolution over the same redshift range (Figure~\ref{fig:dens_evo}).

\item On average, the cumulative number density for the stellar mass ranges studied in this work ($10^{8.48}\leq M_s/\Msun \leq 10^{9.55}$) can be found using,
\begin{equation}
\log n_f = \log n_i-0.10 \Delta z
\end{equation}
for descendants, and,
\begin{equation}
\log n_f = \log n_i+0.12 \Delta z
\end{equation}
for progenitors.
\item  While the evolving cumulative number density is able to track the {\it on average} evolution of galaxies at $3\leq z \leq7$, we find that there is a large scatter of $\sim \pm 0.30$ dex for descendants at $z=$ 3, and $\sim \pm 0.70$ for progenitors at $z=$7 (Figures~\ref{fig:m_evo_d}, \ref{fig:m_evo_p}, \ref{fig:dscatter} \& \ref{fig:pscatter} ).  Individual galaxies demonstrate diverse mass assembly histories due to the physical processes of gas accretion, quenching, mergers and star formation.

\item When evaluating the completeness fraction, we find that only $\sim23\%$ of descendants fall within the original number density bin by $z=3$ (Figure~\ref{fig:comp}).  This low fraction is due to the narrowing of the target mass bin which is caused by the steepening of the CSMF.

\item Increasing the number density bin width at $z=$3 to improve the completeness fraction of $z=$7 descendants to $50\%$ leads the contamination of the original sample by thousands of galaxies which have origins outside of the targeted number density bin.  Thus, significantly reducing the reliability of the evolving cumulative number density method.

\end{itemize}

The technique of utilizing an evolving cut of cumulative number density is a powerful tool in tracking the {\it average} mass growth of galaxies at $3\leq z \leq7$.  This work demonstrates that galaxies in a given number density bin will have very diverse mass assembly histories even if they are of similar stellar mass.  This diversity will reveal itself in the scatter of other physical properties of individual galaxies (i.e. metallicity, age, SFR, star formation history), thus caution should be used when applying this technique to study physical properties other than stellar mass.

\section*{Acknowledgments}

We would like to thank Volker Bromm, Peter Behroozi and Casey Papovich for stimulating conversations.  We are grateful to V. Springel for allowing us to use the original version of {\small GADGET-3} code, on which the \citet{Thompson:14,Jaacks:13} simulations are based. JDJ and SLF would like to acknowledge support from the University of Texas at Austin College of Natural Sciences.  The simulations used in this paper were run and analyzed using pyGadgetReader \citep{pygr} on ‘Blue Waters’ at the National Center for Supercomputing Applications (NCSA). This research is part of the Blue Waters sustained petascale computing project, which is supported by the National Science Foundation (awards OCI-0725070 and ACI-1238993) and the state of Illinois. Blue Waters is a joint effort of the University of Illinois at Urbana-Champaign and NCSA. K.N. acknowledges the partial support by JSPS KAKENHI Grant Number 26247022.



\end{document}